\documentclass[12pt,twoside]{article}
\usepackage{axodraw,epsfig,amsmath,amssymb,latexsym}
\usepackage{float,afterpage,fancybox,array}
\newcommand{\MeV}{{\rm MeV}}
\newcommand{\GeV}{{\rm GeV}}

\newcommand{\fm}{{\rm fm}}

\renewcommand{\Im}{{\rm Im}}
\renewcommand{\Re}{{\rm Re}}

\newcommand{\lsim}{$\raisebox{-0.8ex} {$\stackrel{\textstyle <}{\sim}$}$}

\begin{document}
\title{Nuclear bound states of $\omega$ mesons \thanks{Work supported in part by GSI and BMBF}}
\author{F.~Klingl, T.~Waas\footnote{Present adress: Siemens AG, D-81359 M\"unchen} ~and W.~Weise\\ Physik-Department\\ Technische Universit\"at
M\"unchen\\ D-85747 Garching, Germany}

\maketitle
\begin{abstract}
  Models based on chiral SU(3)$_{\rm L} \, \otimes$
SU(3)$_{\rm R}$  symmetry and vector meson
  dominance suggest an attractive potential for the $\omega$-meson in a nuclear
  medium. We discuss the feasibility of producing nuclear bound states of
  $\omega$-mesons using $(d,^3\hspace{-1mm}He)$ and pion induced reactions on
  selected nuclear targets.
\end{abstract}

\section{\bf Introduction} QCD with massless $u$-, $d$- and $s$-quarks has a chiral SU(3)$_{\rm L} \, \otimes$
SU(3)$_{\rm R}$ symmetry which is known to be spontaneously broken. The scalar
density of quarks develops a non-vanishing ground state expectation value, the
chiral condensate $\langle \bar{q}q \rangle$. The Hellmann-Feynman theorem
suggests that the magnitude of this condensate decreases with
increasing baryon density \cite{1} at a rate determined, to linear order in
density, by the pion-nucleon sigma term. At the density of normal nuclear
matter, $\rho=\rho_0=0.17 \; \fm^{-3}$, the condensate changes from its value at
$\rho=0$ by roughly 30 \%.

Spontaneously broken chiral symmetry is expected to be restored at high
temperatures and possibly at very high baryon densities. At that point the masses of
the ($J^\pi=1^-$) vector mesons should become degenerate with those of their
chiral partners, the ($J^\pi=1^+$) axial vector mesons. One expects that
in-medium vector meson spectra should show traces of this tendency already at
moderate densities. In particular, if vector mesons experience a sufficiently
strong attractive potential in matter, they could form bound states with
ordinary nuclei. 

Such attraction was suggested by in-medium QCD sum rules \cite{4} and by a
simple scaling hypothesis (``Brown-Rho scaling'', \cite{5}). More detailed
analysis \cite{6,7} has revealed that
the $\rho$ meson is probably not a good signal since its large vacuum width of
about 150 MeV will be further magnified by in-medium reactions. In contrast,
the $\omega$ meson appears to be a better candidate \cite{7}. Its small width
in free space (about 8 MeV) might increase by a factor of five or so
 in nuclear matter, but this is still within reasonable limits so that the $\omega$
 meson has a chance to survive as a quasi-particle in the nuclear medium. At
 the same time the calculations suggest that its in-medium mass is shifted
 downward by about 15\% at $\rho=\rho_0$
 \cite{7}, an amount consistent with the original QCD sum rule analysis
 \cite{4}. While much of the recent activities have focused on possible
 in-medium modifications of vector mesons spectra in the context of
 ultra-relativistic heavy-ion collisions, it is certainly useful to explore
 such effects under less extreme, better controled conditions: hence the quest
 for possible $\omega$ meson (quasi-) bound states in ordinary nuclei. Similar
 questions have been raised in refs.~\cite{8,9}.

Our paper proceeds as follows. In section 2 we present an update of the $\omega
N$ interaction and the in-medium spectrum of the $\omega$ meson. Section 3
focuses on the $\omega$-nuclear optical potential and its quasi-bound states. In
section 4 we explore possible mechanisms for producing such states: the 
$(d,^3\hspace{-1mm}He)$ reaction and pion-induced $\omega$
production. Concluding remarks follow in section 5.

\section{\bf Update on the $\omega$ meson self-energy \protect \newline in nuclear matter} 

The framework of our discussions is an effective Lagrangian, based on chiral
SU(3)$_{\rm L} \, \otimes$ SU(3)$_{\rm R}$ symmetry, which incorporates the
octet of pseudoscalar Goldstone bosons (pions, kaons, ...), baryons and vector
mesons \cite{10,7}. It specifically includes anomalous couplings of the
$\omega$ meson related to the Wess-Zumino term.

In ref.~\cite{7} we have calculated in-medium spectra of $\rho, \; \omega$ and
$\phi$ mesons using this framework. For the $\omega$ meson, the basic input is
the $\omega$-nucleon amplitude in free space, $T_{\omega N}(E)$, as a function
of energy $E$ at zero three-momentum, $\vec{q}=0$. Restricting the discussion
to $\vec{q}=0$ is justified since $\omega$ mesons bound to nuclei will have
small average momenta, $\langle \vec{q}^{\,2}\rangle \ll m_\omega^2$. 

In comparison with our previous results for $T_{\omega
  N}$ presented in ref.~\cite{7}, improvements have recently been made by more
rigorous inclusion of constraints from the measured $\pi^- p \to \omega n$
cross section \cite{11}. We have also examined uncertainties arising from the
incompletely known (off -shell) behaviour of $T_{\omega N}(E)$ at large $E$.
We have performed further calculations of $T_{\omega N}$ as specified in
ref.~\cite{7} where all the technical details can be found, but using two
different input sets, A and B \cite{12}. These sets differ in their treatment
of $\omega N \to \pi \pi N$ processes involving intermediate $\rho N$ states,
as illustrated in Fig.~\ref{diag5}. Set A includes the box diagrams,
Figs.~\ref{diag5}a and b. Set B inorporates in addition the diagram,
Fig.~\ref{diag5}c, in which the $\rho N$ interaction vertex has large tensor coupling ($\kappa_\rho
\simeq6$). Fig.~\ref{fig1a}a shows the real and imaginary parts of $T_{\omega
N}(E,\,\vec{q}=0)$ calculated with sets A and B, respectively. The imaginary
parts are evalutated as described in ref.~\cite{7} (choosing the sign of the
$\omega \rho \pi$ coupling to be consistent with the constraints from $\pi^- p
\to \omega n$), and the real parts are obtained using the dispersion relation
(\ref{eq2}). The differences between the solid and dashed curves (sets A vs. B)
give a rough impression of the model dependence of $T_{\omega N}$ at high energy
due to uncertainties in the $\omega N \leftrightarrow \rho N$ coupled channel
dynamics.

\begin{figure}
\vspace*{-2cm}
\begin{center}
\unitlength0.32mm
\begin{picture}(400,150)(0,0)
\SetWidth{1}    
\SetScale{0.9} 
\ArrowLine(10,15)(30,15)
\ArrowLine(30,15)(70,15)
\ArrowLine(70,15)(90,15)
\SetWidth{3}    
\Line(30,15)(30,55)
\Line(70,15)(70,55)
\SetWidth{1}    
\DashLine(30,55)(70,55){4}
\DashCArc(50,15)(20,0,180){4}
\ZigZag(10,75)(30,55){3}{3}
\ZigZag(70,55)(90,75){3}{3}
\Text(50,0)[]{(a)}
\Text(20,40)[]{$\rho$}
\Text(80,40)[]{$\rho$}
\Text(50,65)[]{$\pi$}
\Text(50,40)[]{$\pi$}
\Text(20,76)[]{$\omega$}
\Text(0,10)[]{$N$}
\Vertex(30,15){2}
\Vertex(70,15){2}
\Vertex(30,55){2}
\Vertex(70,55){2}

\ArrowLine(160,15)(180,15)
\ArrowLine(180,15)(220,15)
\ArrowLine(220,15)(240,15)
\SetWidth{3}
\Line(180,55)(220,55)
\SetWidth{1}
\DashLine(180,15)(180,55){4}
\DashLine(220,15)(220,55){4}
\ZigZag(160,75)(180,55){3}{3}
\ZigZag(220,55)(240,75){3}{3}
\Text(205,0)[]{(b)}
\Text(150,10)[]{$N$}
\Text(205,75)[]{$\rho$}
\Text(170,35)[]{$\pi$}
\Text(230,35)[]{$\pi$}
\Text(150,78)[]{$\omega$}
\Vertex(180,15){2}
\Vertex(220,15){2}
\Vertex(180,55){2}
\Vertex(220,55){2}

\ArrowLine(310,15)(330,15)
\ArrowLine(330,15)(420,15)
\ArrowLine(420,15)(440,15)
\ZigZag(310,35)(330,15){3}{3}
\ZigZag(420,15)(440,35){3}{3}
\SetWidth{3}
\CArc(375,15)(25,0,180)
\SetWidth{3}
\Text(375,0)[]{(c)}
\Text(375,50)[]{$\rho$}
\Text(325,40)[]{$\omega$}
\Text(300,10)[]{$N$}
\Vertex(330,15){2}
\Vertex(420,15){2}
\Vertex(350,15){2}
\Vertex(400,15){2}

\end{picture}
\end{center}
\caption{\label{diag5} \protect \newline
Contributions to $T_{\omega N}$ involving the $\omega N \to \pi \pi N$
reaction channel. The thin
solid line represents the nucleon. The $\rho$ meson propagators in diagrams (b)
and (c) include $\rho \to \pi \pi$ decay. Set A incorporates diagrams (a) and
(b). Set B sums up all three diagrams, also including mixed terms between (b)
and (c).}
\end{figure}

We note that $T_{\omega N}$ is strongly energy dependent, its real part changing
from repulsion at $E=0$ to attraction at $E\sim m_\omega$. The imaginary part
rises quickly at energies $E>m_\omega$. It represents the sum of the $\omega N
\to \pi N,\; \pi \pi N$, etc. reaction channels (see Fig.~\ref{fig1a}b). The $\omega N \to
\pi \pi N$ channel is actually the prominent one for $E \gtrsim 0.7 \; \GeV$,
while the relatively small $\omega N \to \pi N$ reaction width determines $\Im
T_{\omega N}$ at $E \lesssim 0.6 \, \GeV$. 

Our calculated effective scattering length (with set A),
\begin{equation}
  \label{eq1}
  a_{\omega N} = \frac{M_N}{4 \pi (M_N+m_\omega)} T_{\omega N}
  (E=m_\omega)=(1.6+i0.3) \, \fm,
\end{equation}
suggests a substantial downward shift of the in-medium $\omega$ mass and a
moderately increasing width in matter\footnote{Our updated values in eq.(\protect \ref{eq1}) are smaller than those in ref.~\protect\cite{7} where the $\omega N \to
\pi N$ reaction width had been overestimated. This has now been
corrected.}. This downward shift comes in large part from $2 \pi$-exchange
processes such as Fig.~1(b) which, at the same time, have large inelasticities
from $\omega N \to \rho N \to \pi \pi N$.

Remaining uncertainties concern the constant $T_{\omega N}(E=0)$ which
enters into the dispersion relation
\begin{equation}
  \label{eq2}
 \Re \, T_{\omega N} (E) = T_{\omega N}(0)+ \frac{E^2}{\pi}
  {\cal P} \int_{0}^\infty du^2 \frac{\Im \, T_{\omega N} (u)}{u^2
  (u^2-E^2)},
\end{equation}
used to derive $\Re \, T_{\omega N} $ from the calculated $\Im \, T_{\omega N}$.
Vector Meson Dominance (VMD) connects $T_{\omega N}(0)$ with the Thomson limit of
the Compton scattering amplitude, so that $T_{\omega N} (0)=-\left(\frac{3 g}{2} \right)^2 M_N^{-1}$ with the nucleon mass $M_N$ and the universal
vector coupling constant $g=m_\rho/(\sqrt{2} f_\pi) \simeq 5.9$. Deviations from
VMD could arise from additional contact terms proportional to $\bar{N}N
\omega_\mu \omega^\mu$ in the effective Lagrangian, so that the constant
$T_{\omega N } (0)$ remains basically unconstrained.

Nevertheless, a first order estimate for the
$\omega$ meson mass shift in nuclear matter gives, using eq.(\ref{eq1}):
\begin{equation}
  \label{eq3}
\frac{\Delta m_\omega (\rho)}{m_\omega}=-\frac{2 \pi \rho}{m_\omega^2}
\left(1+\frac{m_\omega}{M_N} \right) \Re\; a_{\omega N} \simeq -0.2 \frac{\rho}{\rho_0},
\end{equation}
which is consistent with the in-medium QCD sum rule analysis \cite{4,7}. The in-medium QCD
sum rule does not have quantitative predictive power because of uncertainties related to the treatment of the density
dependence of four-quark condensates, but it is still useful for
orientation. It provides a model independent consistency test for the first
moment of the $\omega$ meson spectral distribution \cite{13}, in vacuum as well
as in nuclear matter. 

The detailed evaluation \cite{7} of the in-medium $\omega$ meson spectrum
involves the self-energy tensor, $\Pi^{\mu \nu
  }(E,\vec{q}\,; \rho)$, of the $\omega$ meson in nuclear matter at density
  $\rho$. For vanishing momentum $\vec{q}$ its longitudinal and transverse
  parts $\Pi^{(L, T)}(\omega,\vec{q};\rho)$  coincide, and we introduce 
\begin{equation}
  \label{eq4}
  \Pi(E;\rho)\equiv -\frac{1}{3} \Pi^\mu_\mu (E,\vec{q}=0;\rho)=\Pi^{(T)}(E,\vec{q}=0;\rho)=\Pi^{(L)}(E,\vec{q}=0;\rho).
\end{equation}
\begin{figure}[ht]
\vspace*{-0.9 cm}
\unitlength1mm
\begin{picture}(150,75)
\put(0,0){\makebox{\epsfig{file=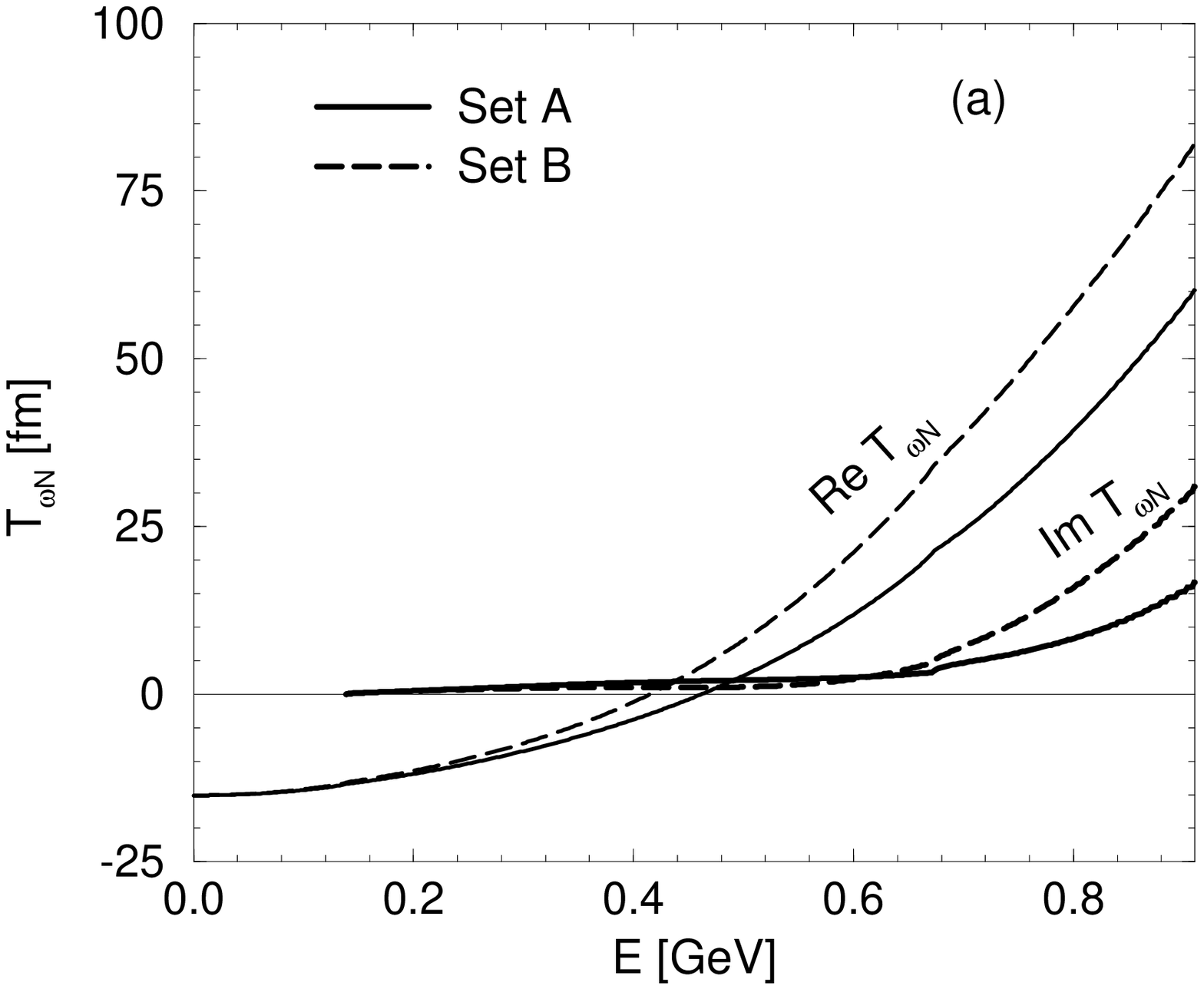,width=84mm}}}
\put(80,0){\makebox{\epsfig{file=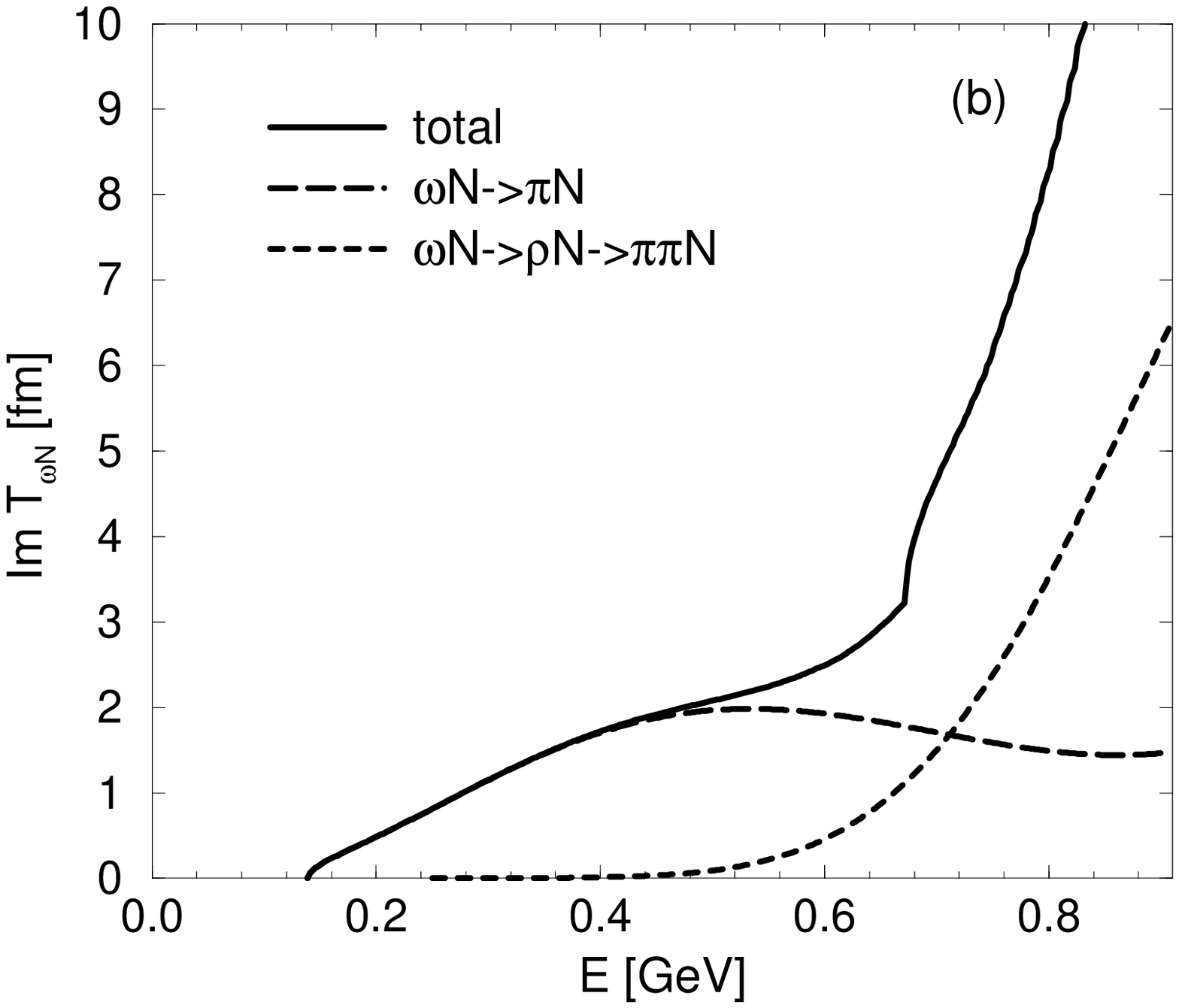,width=84mm}}}
\end{picture}
\vspace*{-4mm}
\caption{\label{fig1a} a) Real and imaginary parts of the $\omega N$ amplitude
$T_{\omega N} (E,\vec{q}=0)$ calculated with set A (solid) and set B
(dashed). For further details see ref.~\cite{7}. \protect \newline b)
Contributions to $\Im T_{\omega N}(E,\vec{q}=0)$ from $\omega N \to \pi N$ and
$\omega N \to \pi \pi N$ reaction channels together with the total result
(solid line) for Set A.
}
\end{figure}
At $\rho=0$, the imaginary part of $\Pi$ gives the vacuum decay width of the
$\omega$ meson, and its real part is absorbed into the physical free mass
$m_\omega$. The density dependent part of $\Pi$ involves the $\omega$-nuclear
interaction which is treated to leading order in the density $\rho$. The
in-medium propagator for the $\omega$ meson ``at rest'',
$D_\omega=-(E^2-m_\omega^2-\Pi)^{-1}$, reduces to:
\begin{equation}
  \label{eq5}
  D_\omega(E,\vec{q}=0;\rho)=\frac{-1}{E^2-m_\omega^2+i E\, \Gamma_\omega^{(0)}(E)
  +\rho \, T_{\omega N}(E)},
\end{equation}
with $\Gamma_\omega^{(0)}=8.4 \, \MeV$ at $E=m_\omega$ and the complex $\omega N$
amplitude $T_{\omega N}$ as given in Fig.~\ref{fig1a}(a). Note that $
\Gamma_\omega^{(0)}(E)$ represents the $\omega \to \pi^+ \pi^0\pi^-$ decay and
grows even faster with energy $E$ than the $3 \pi$ phase space.

\begin{figure}[ht]
\vspace*{-0.9 cm}
\unitlength1mm
\begin{picture}(160,75)
\put(40,0){\makebox{\epsfig{file=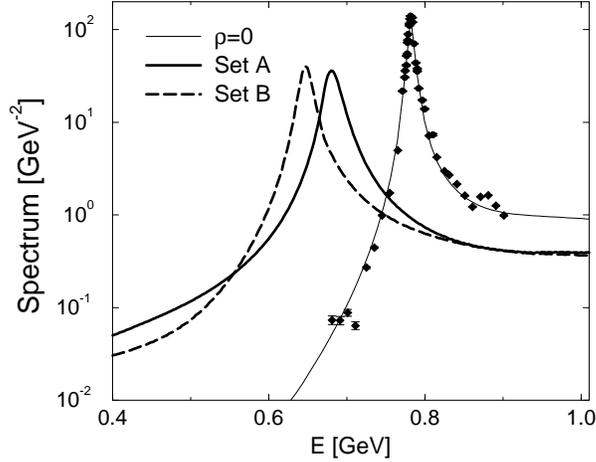,width=84mm}}}
\end{picture}
\vspace*{-4mm}
\caption{\label{fig1b} a) Spectral distribution $\Im \, D_\omega(E,\vec{q}=0)$
of the $\omega$ meson in vacuum (thin line, with data) and in nuclear matter at
density $\rho=\rho_0=0.17\, \fm^{-3}$, for set A (solid) and set B
(dashed). The data points correspond to $e^+e^- \to \omega \to 3 \pi$ in vacuum
\cite{14}.} 
\end{figure}

The spectral distribution $\Im D_\omega$ at nuclear matter density $\rho=\rho_0=0.17 \;
\fm^{-3}$ is shown in Fig.~\ref{fig1b} in comparison with the vacuum ($\rho=0$)
spectrum. The solid and dashed curves at $\rho=\rho_0$ reflect the model
dependence in the calculations. Despite such uncertainties, the following
features are clearly visible:
\begin{itemize}
\item The $\omega$ meson (unlike the $\rho$ meson) is expected to persist as a
quasi-particle in nuclear matter when it has a small momentum relative
to the surrounding medium.
\item The in-medium $\omega$ meson mass at $\rho=\rho_0$ is shifted downward by
about 15 \% from the free $m_\omega$.
\item The predicted in-medium $\omega$ meson decay width (at resonance) is
$\Gamma_{\rm eff}(\rho=\rho_0) \simeq 40$ MeV, about five times as large as the
free width, but still about an order of magnitude smaller than $m_\omega$.
\end{itemize}

We repeat that our calculated $\omega$ spectrum in matter is perfectly
consistent with the QCD sum rule analysis \cite{4,7}. In fact, the first moment
of this spectral distribution (which does {\cal not} depend on the four-quark
condensate but only on the gluon and quark condensates together with the first
moment of the quark distribution known from deep inelastic lepton-nucleon
scattering) follows precisely the trend expected from QCD sum rules \cite{13}.

We use our predicted
spectrum as basis for the discussion of possible $\omega$ meson quasi-bound
states. The term ``quasi-bound'' should indicate that such bound states, if
existent, will have short lifetimes due to their hadronic decays, but the
widths are not overwhelmingly large according to our calculated $\Im T_{\omega
N}$.
\newpage

\section{ The $\omega$-nuclear potential and quasi-bound states}

In the local density approximation, the $\omega$ meson self-energy (\ref{eq4})
can be translated into a complex, energy dependent local potential:
\begin{align}
  \label{eq6}
  \Pi(E, \vec{r}\,) &\equiv  2 E U(E,\vec{r}\,)\\ \nonumber &= - \Re T_{\omega
  N}(E) \rho(\vec{r}\,)-i [ E \Gamma_\omega^{(0)}(E)+\Im T_{\omega N}(E)\, \rho(\vec{r}\,)].
\end{align}
Notably, $\Re U$ can reach values of order $-100$ MeV in the center of nuclei
and at the energies in question. Non-local terms proportional to  $\vec{\nabla}
\rho(\vec{r}\,) \cdot \vec{\nabla}$ are expected to be less important and will be
dropped in the present survey calculation. Given that the longitudinal and
transverse parts of the self-energy are the same in the limit $\vec{q}\to0$,
all four vector components of the $\omega$ meson field are described by a
single wave function $\phi(\vec{r\,})$ which satisfies the Klein-Gordon equation
\begin{equation}
  \label{eq7}
   \left[E^2+\vec{\nabla}^2-m_\omega^2-\Pi(E, \vec{r}\,) \right] \phi (\vec{r}\,)=0.
\end{equation}
Quasi-bound $\omega$-nuclear states are defined by self-consistent solutions of
eq.~(\ref{eq7}) at complex energies $E_\lambda$ for which $\Re E_\lambda< m_\omega$. We
introduce
\begin{equation}
  \label{eq8}
  {\cal E}_\lambda =  \Re E_\lambda-  m_\omega
\end{equation}
and refer to $-{\cal E}_\lambda$ as the binding energies of these states. Their
total decay widths are given by 
\begin{equation}
  \label{eq9}
  \Gamma_\lambda=-2\,  \Im E_\lambda.
\end{equation}

\begin{table}
\begin{center}
\begin{tabular}{|cc||c|c|c|} \hline
     && \multicolumn{3}{c|}{$({\cal E}_{nl}, \Gamma_{nl})$ \, [MeV]}   \\
   nucleus& n& l=0&  l=1&  l=2  \\ \hline \hline 
  $^6_\omega He$:& (1) & (-49,32)&(-19,26)     & -\\ \hline   
   $^{11}_\omega B$:& (1)& (-66,38)& (-40,34) &(-13,30) \\
   & (2)& (-14,26)&-&- \\    \hline   
   & (1)&(-83,42)& (-64,40)&(-45,40) \\
   $^{26}_\omega Mg$:   & (2)& (-42,38)&(-21,36) &(-1,28) \\
   &(3)& (-2,24)&-&- \\ \hline
   & (1)&(-87,42) &(-74,42)& (-57,42) \\
   $^{39}_\omega K$:&(2)& (-54,40)&(-36,40)&(-17,36) \\
   &(3)&( -16,34)&-&- \\ \hline 
\end{tabular}
\end{center}
\caption{\label{tab1} Complex eigenenergies,  $E_{nl}=m_\omega+{\cal
E}_{nl}-\frac{i}{2} \Gamma_{nl}$, obtained by solving the Klein-Gordon equation
(\ref{eq7}) with the potential (\ref{eq6}) and density (\ref{eq10}) using set A
as input.}
\end{table}

For the nuclear density distributions we use 
two-parameter Fermi profiles
\begin{equation}
  \label{eq10}
  \rho(r)= \frac{\cal N}{1+e^{(r-R)/a}},
\end{equation}
with $R=(1.18 A^{\frac{1}{3}}-0.5)$ fm and $a=0.5$ fm. The normalization constant
${\cal
  N}$ is determined by $\int d^3r \rho=A$.

In table 1 we summarize the complex eigenenergies found by solving
eq.(\ref{eq7}) for selected (somewhat unusual) light and medium weight nuclei. The
reason for their choice is that these nuclei can be formed in
$(d,^3\hspace{-1mm}He)$ reactions as in the GSI proposal \cite{8}. The
calculations have been done with spherically symmetric density distributions (\ref{eq10}),
so that the quasi-bound $\omega$ meson states are labeled by their principal
quantum number and their angular momentum, $\lambda=n,l$.

One observes that the attractive potential (\ref{eq6}) supports quasi-bound
omega meson states even for light nuclei. The examples given in table 1 should
be representative for a wider class of nuclei, and we do not expect that
detailed nuclear structure effects such as deformations change this picture
qualitatively. These quasi-bound states are of course short-lived, with full
widths that are generally of similar magnitude as their binding
energies. Unless these widths are grossly underestimated, one should be able to
see spectral strength at energies below the free $\omega$ mass, even if these
states may not be resolved individually.

Our predictions in table 1 are in qualitative agreement with the findings of
ref.~\cite{9}.  They differ at a quantitative level since our $\omega$-nuclear
potential has the important feature that it is strongly energy dependent. The
strength of the attractive real part of our potential tends to grow with
increasing energy in the vicinity of $E \simeq m_\omega$, so that it actually
supports more quasi-bound states than the potential of ref.~\cite{9}. The
energies of the lowest $(1s)$ states are nevertheless strikingly similar in
both calculations. The primary point of our approach is that we can offer an estimate, based on explicit calculations, for the widths of such
states with realistic constraints from the prominent $\omega N$ reaction
channels.

Our intermediate conclusion is the following: if the $\omega$ meson mass
experiences a downward shift in nuclear matter by $10 -15\%$ at $\rho=\rho_0$,
then this effect should be visible already in ordinary, even light nuclei by
the appearance of quasi-bound states.

\section{Production of $\omega$-nuclear states}
\subsection{\label{4.1} Transfer reaction}
Hayano et al.~\cite{8} suggested to produce bound $\omega$ states through
transfer reactions of the type $d+A \to\; ^3\!He+\omega(A-1)$. Here the
incoming deuteron with momentum $\vec{p}_d$ picks up a bound proton in the
target nucleus $A$ to form an outgoing $^3\!He$ together with an $\omega$ meson
attached to the residual $(A-1)$ nucleus. The final $^3\!He$ energy $E_{He}$ and momentum $\vec{p}_{He}$ is
measured in forward direction. The basic idea is to minimize the momentum
$\vec{q}=\vec{p}_{He}-\vec{p}_d$ transfered to the residual system with its
bound $\omega$ meson. Energy and momentum conservation implies that such
conditions would ideally be met with a deuteron beam of kinetic energy $T_d
\simeq 10$ GeV to produce the $\omega$ meson at rest. Even for the lower
deuteron beam energies available at GSI ($T_d \simeq 4$ GeV) the situation
would still be favourable since the momentum transfers would be around $q
\sim0.2$ GeV, well within the range of the typical of nucleon Fermi momentum 
in the nucleus.
\begin{figure}[ht]
\begin{center}
\unitlength1mm
\begin{picture}(70,70)
\put(0,0){\epsfig{file=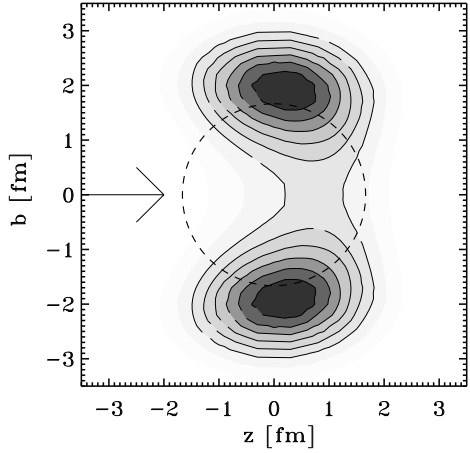,width=90mm}}
\end{picture}
\end{center}
\vspace{-6mm}
\caption{ \label{fig13}   Contour plot of $D(z,b) \rho(r)$ indicating the
reaction zone for the $(d,^3\hspace{-1mm}He)$ process on a typical light target
nucleus. The dashed circle indicates the r.m.s radius of the target. }
\end{figure}

The framework of our calculation is the distorted wave impulse
approximation. The $(d,^3\hspace{-1mm}He)$ differential cross section for $\omega$ meson
production is written
\begin{equation}
  \label{eq11}
 \left(\frac{d^2 \sigma_{d+A \to ^3\!He+\omega(A-1) }}{d E \, d \Omega}
 \right)^{\rm lab}_{\theta=0} = \left(\frac{d \sigma_{d+p \to ^3\!He +\omega} }{ d \Omega}
 \right)^{\rm lab}_{\theta=0} S(E).
\end{equation}
The free cross section $(d \sigma_{d+p \to ^3\!He +\omega} /d \Omega
)$ has been measured at SATURNE \cite{15}. Its value in the lab frame at $T_d
\simeq 4
\, GeV$ is approximately $0.5\, {\rm \mu b/sr}$.

The task is then to calculate the response
function $S(E)$ which describes the removal of a proton from the target nucleus
and the formation of the final state, with the $\omega$ meson attached to the
residual nucleus, and it includes the distorted waves of the incoming deuteron
and outgoing $^3\hspace{-1mm}He$. The energy variable is $E=T_d+M_d-E_{He}$
which roughly equals $E_\omega+B_p$, the energy of the produced $\omega$ meson
plus the binding energy of the removed proton. 

The calculation of $S(E)$ is performed using a nuclear shell model with
realistic single particle wave functions for the bound proton. Its generic form
is \cite{16} 
\begin{eqnarray}
  \label{eq12}
  S(E)= \frac{1}{\pi} \sum_f \;{\cal F}^\dagger (dp\to ^3\!He) \;
  G_{\omega} (E-B_p) \;
 {\cal F}(dp \to ^3\!He),
\end{eqnarray}
where the sum is taken over all possible final states that can be reached at
given energy transfer $E$.
Here
\begin{equation}
  \label{eq13}
   G_\omega (E)=\sum_\lambda |\phi_\lambda \rangle \frac{-2
   E}{E^2-E_\lambda^2+i \epsilon} \langle \phi_\lambda|
\end{equation}
is the $\omega$ meson Green function related to the Klein-Gordon equation
(\ref{eq7}). The amplitude ${\cal F}$ involves the bound proton wave function
\begin{equation}
  \label{eq14}  
  \Psi_{j_p,l_p} (\vec{r}\,)= \psi_{j_p l_p}(r) [Y_{l_p}(\hat{r}) \chi_{\frac{1}{2}}]^{[j_p]}.   
\end{equation}
It also involves the distorted waves, $\Psi_i(\vec{p}_{d},\, \vec{r}\,)$ of the
incoming deuteron and $\Psi_f(\vec{p}_{He},\,\vec{r}\,)$ of the outgoing
$^3\!He$. Given the high energies of these particle, we can use the eikonal
approximation and write (with their momenta $\vec{p}_d$ and $\vec{p}_{He}$
chosen along the $z$-axis):
\begin{equation}
  \label{eq15}
  \Psi_f^\dagger (\vec{p}_{He},\vec{r}\,) \, \Psi_i(\vec{p}_d,\vec{r}\,) = e^{i (p_d-p_{He}) z} \, D(\vec{r}).
\end{equation}
We introduce $z=r cos \Theta$ and the impact parameter $b=\sqrt{x^2+y^2}=r sin
\Theta$ and write the distortion factor $D(\vec{r}\,)$ as:
\begin{equation}
  \label{eq16}
  D(\vec{r}\,)= \exp{\left[-\frac{\sigma_{d N}}{2} \int_{-\infty}^{z} dz'  \,
  \rho(z',b) - \frac{\sigma_{^3\!He N}}{2} \int_{z}^{\infty} dz' \, \rho(z',b)
  \right]} .
\end{equation}
We use $\sigma_{d N} \simeq 83 \;{\rm
  mb}$ and $\sigma_{^3\!He N }\simeq 126 \;{\rm mb}$ for the
deuteron-nucleon and $^3\!He$-nucleon cross section at typical beam
energies around 4 GeV. 

Carrying out all necessary spin and angular momentum sums (including the
polarization degrees of freedom of the $\omega$ meson) the response function
reduces, after some algebra, to the following expression:
\begin{eqnarray}
  \label{eq17}
&& S(E)= \sum_{j_p,l_p} \sum_{l,L} N_p \frac{2 l+1}{4 \pi^2} (l_p 0
  l 0|L0) \cdot \\ && \Im \int_0^\infty dr' {r'}^2\,w_L^*(r')\;\psi^*_{j_p l_p}(r')
 \int_0^\infty dr  r^2  \, w_L(r) \; \psi_{j_p l_p}(r)\,{g}_{l}(E-B_p,r',r),\nonumber
\end{eqnarray}
where $\psi_{j_p l_p}$ is the proton radial wave function specified in
eq.(\ref{eq14}), $N_p$ denotes the number of protons in the specific orbital,
and
\begin{equation}
  \label{eq18}
  g_{l}(E;r',r)= 2i \,E\, k \,u_{l}(k,r_<)\, v^*_{l}(k,r_>)
\end{equation}
is the remaining radial Green function of the $\omega$ meson for given
angular momentum $l$, with  $r\equiv|\vec{r}\,|$ and $ r'\equiv |\vec{r}\,'|$. It is written in terms of the regular solution 
$u_{l}$ and the outgoing solution $v^*_{l}$ of the 
Klein-Gordon equation and applies both for continuum states and for quasi-bound
states. The wave number is $k=\sqrt{E^2-m_\omega^2}$, and bound states are
expressed by analytic continuation to imaginary $k$. Furthermore,
\begin{equation}
  w_L(r)= \int_{-1}^{1}dcos \Theta \, e^{i (p_d-p_{He}) r cos \Theta
  }\,D(z(\Theta),b(\Theta))\, P_L(cos \Theta),
\end{equation}
with Legendre polynomials $P_L$. Note that the $(d,^3\hspace{-1mm}He)$ transfer
reaction takes place predominantly at the edges of the nuclear target, as
shown in Fig.4. The distortion factor causes a suppression of the cross section
by typically two orders of magnitude.

Substituting the response function (\ref{eq17}) into the cross section
(\ref{eq11}) we can now study the effects of $\omega$-nuclear binding on the
missing energy spectrum measured in the transfer reaction. We choose the case
$^7\!Li(d,^3\hspace{-1mm}He)^6_\omega\hspace{-1mm}He$ as an example, mainly for
simplicity of its proton orbital structure and since this choice is referred to
in the GSI proposal \cite{8}. The result is shown in Fig.5. The differential
cross section is small, but a systematic downward shift of strength as compared
to quasi-free $\omega$ meson production should be nevertheless visible, even in
light nuclei. The effect should be further enhanced in heavier nuclei. The
increased $\omega$ meson width in the nuclear environment prohibits a separate
identification of isolated bound states, but the overall downward shift of the
$^3\!He$ missing energy spectrum below the free $\omega$ meson threshold would
indicate the presence of quasi-bound states if they exist. A flat background,
not shown here, is expected to come primarily from $\rho$ meson production. The
model dependence of the input $\omega N$ amplitude is again illustrated by
showing results for sets A and B discussed previously. The qualitative
conclusions are evidently not much influenced by the difference between these
choices.
\begin{figure}[ht]
\unitlength1mm
\begin{picture}(140,68)
\put(0,0){\makebox{\epsfig{file=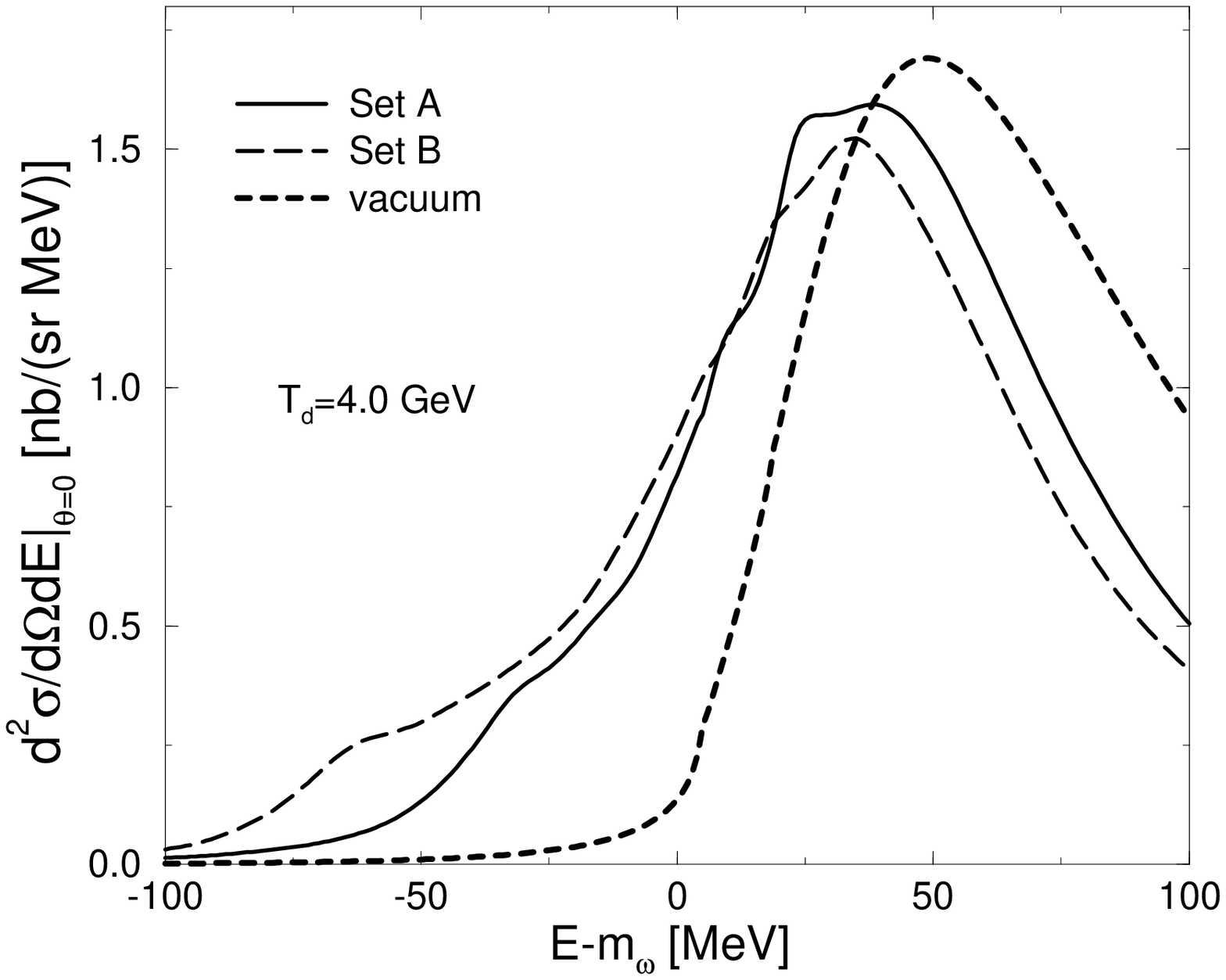,width=88mm}}}
\put(78,0){\makebox{\epsfig{file=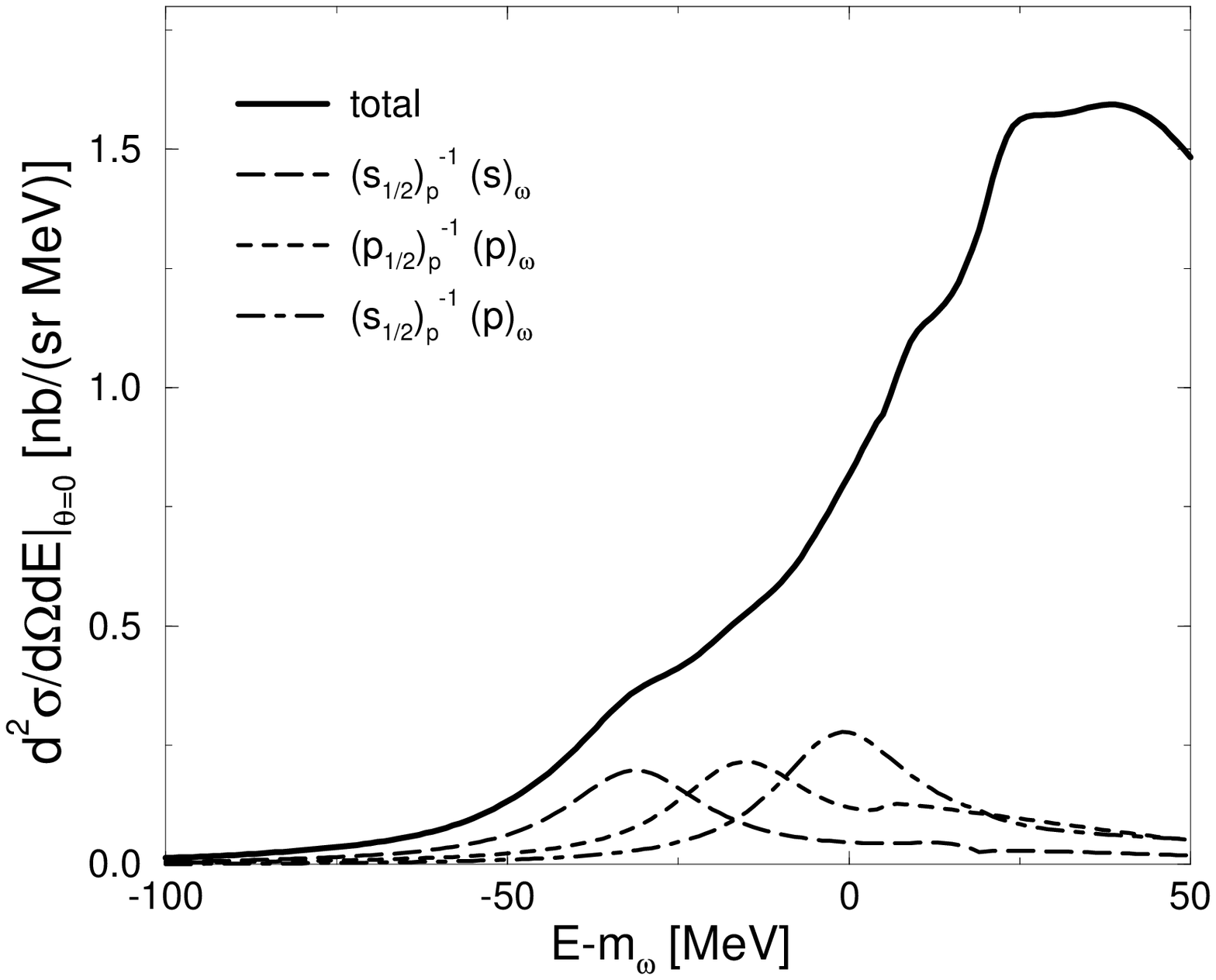,width=88mm}}}
\end{picture}
\vspace*{-9mm}
\caption{\label{espec} Missing energy spectra of the reaction
$d+A\to ^3\hspace{-2mm}He+\omega(A-1)$ at a deuteron kinetic energy $T_d=4$ GeV in case of A=7 (Lithium): \protect \newline
a) Cross sections calculated with $\omega$-nuclear potentials, Set A 
  (solid line) and Set B (long dashed line), together with the ``vacuum''
  result (quasi--free $\omega$ production, short dashed).
\protect \newline 
b) Separate contributions from transitions of protons of a
specified shell (subscript $p$) to bound states of $\omega$ mesons  (subscript
$\omega$). The total spectrum corresponds to Set A.}
\end{figure}
Fig.~5b shows the decompostion of the quasi-bound spectrum into separate ($s$-
and $p$-state) orbitals of the $\omega$ meson coupled to different proton-hole
states.

\subsection{Pion-induced $\omega$ production}
Another interesting option is the $\pi^-p \to \omega n  $ reaction in nuclei
\cite{18}. For a neutron emitted in forward direction, a pion with lab. kinetic
energy $T_\pi \simeq 2.5\, \GeV$ incident on  a proton at rest produces a zero
momentum $\omega$ meson. For an $\omega$ meson bound to a nucleus, $T_\pi \simeq
(1.5-2) \, \GeV$ may be sufficient. Consider the case in which the
$\omega$ meson energy $E_\omega$ is measured by observing its decay $\omega \to
e^+ e^-$, whereas the emitted neutron remains unobserved. We wish to estimate
the differential cross section for this process. 

The $z$-axis of the lab. frame is chosen to coincide with the incoming pion
momentum. The cross section is written
\begin{figure}[ht]
\begin{center}
\unitlength1mm
\begin{picture}(70,70)
\put(0,0){\makebox{\epsfig{file=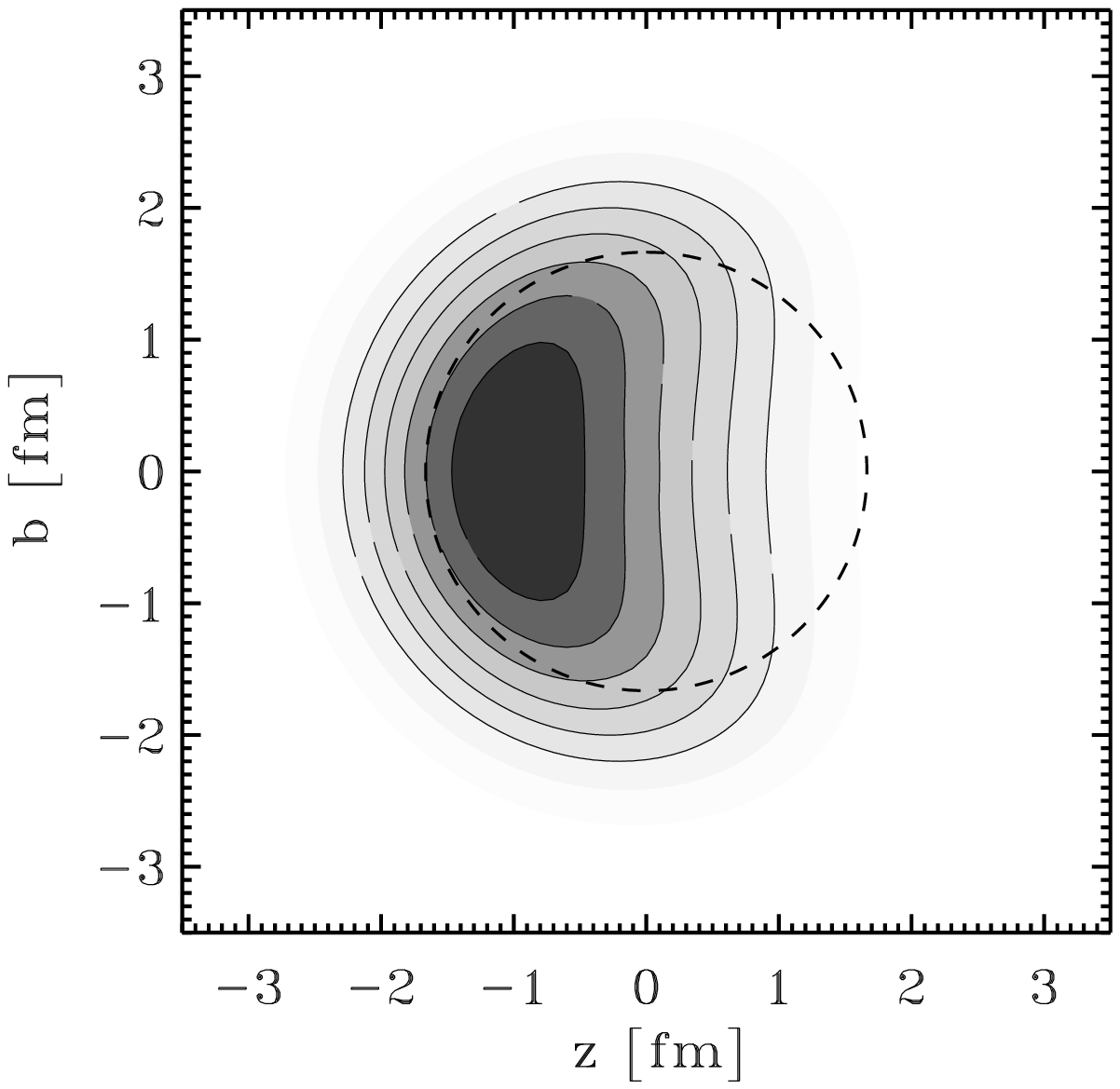,width=90mm}}}
\end{picture}
\vspace{-6mm}
\caption{\label{fig13b}  Contour plot of  $D(z,b) \rho(r)$ indicating the
reaction zone for the $\pi^- A \to \omega (A-1)$ process on a typical light
target nucleus. The dashed curve indicates the r.m.s. radius of the target. }
\end{center}
\end{figure}

\begin{figure}
\begin{center}
\unitlength1mm
\begin{picture}(80,60)
\put(0,0){\epsfig{file=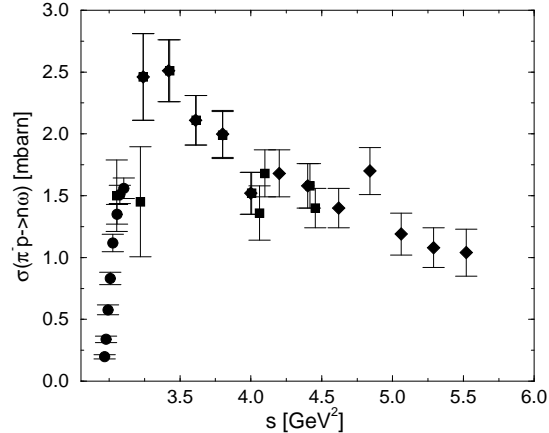,width=80mm}}
\end{picture}
\end{center}
\vspace{-9mm}
\caption{\label{fig14b} The free cross section of the reaction $\pi^-+p
\to n +\omega$. The data are from ref.\protect \cite{15}.}
\end{figure}  
\begin{equation}
  \label{eq20}
  \frac{d \sigma}{d E_\omega} = \frac{\Gamma_{\omega \to e^+
  e^-}}{\Gamma_{\rm tot}}  \int d \Omega_n \frac{d \sigma_{\pi^- p \to
  \omega n}^{\rm Lab}}{d \Omega_n}
  \cdot S(E_\omega,\Omega_n),
\end{equation}
where the integral is taken over the solid angle of the emitted neutron. The
factor $\Gamma_{\omega \to e^+ e^-}/\Gamma_{\rm tot}$ is the branching ratio
for the dilepton decay channel of the $\omega$ meson, $7.15 \cdot
10^{-5}$ for the free $\omega$ and possibly up to an order of magnitude smaller for the
in-medium decay. The c.m. cross section for $\pi^- p \to n \omega$ in free
space (see Fig.~7) is
almost isotropic and in the range 1-2 mb \cite{17}. One just needs to transform
this quantity to the lab frame.

It remains to calculate the response function $S(E_\omega,\Omega_n)$ which now
depends on the neutron emission angle. We follow similar steps as in the
previous section \ref{4.1}. The (undetected) energetic neutron in the final
state is treated as a plane wave $\Psi_f(\vec{p}_n,\,\vec{r}\,)=\exp(i\vec{p}_n\cdot
\vec{r}\,)$. The distorted wave of the incident pion is written 

\begin{equation}
  \label{eq21}
\Psi_\pi (p_\pi,\vec{r}\,) = e^{i
  p_\pi z}  D(\vec{r}\,),
\end{equation}
with the distortion factor
\begin{equation}
  \label{eq22}
  D(\vec{r}\,)= \exp{\left[-\frac{\sigma_{\pi N}}{2} \int_{-\infty}^{z} dz' \,
  \rho(z',b)\right]} \equiv D(z,b).
\end{equation}
We use $\sigma_{\pi N} \simeq 35$ mb.

Note that the reaction zone is now concentrated in the front hemisphere of the target nucleus (see Fig.~6). The final result for
the $\pi^- A \to \omega (A-1)+n$ response function becomes:
\begin{eqnarray}
  \label{eq23}
    && S(E_\omega,\Omega_n)= \sum_{j_p,l_p} \sum_{l,L,M} N_p\, \frac{2 l+1}{4
    \pi^2}\,(l_p0l  0|L0) \cdot\\
  &&\Im  \int_0^\infty dr' {r'}^2 \,w_{L M}^*(r',\theta_n) \,
  \psi^*_{j_p l_p}(r') \int_0^\infty dr  r^2\, w_{L M}(r,\theta_n) \,\psi_{j_p
  l_p}(r) \, g_l(E_\omega,r',r),\nonumber
\end{eqnarray}
\begin{figure}[ht]
\unitlength1mm
\begin{center}
\begin{picture}(80,68)
\put(0,0){\epsfig{file=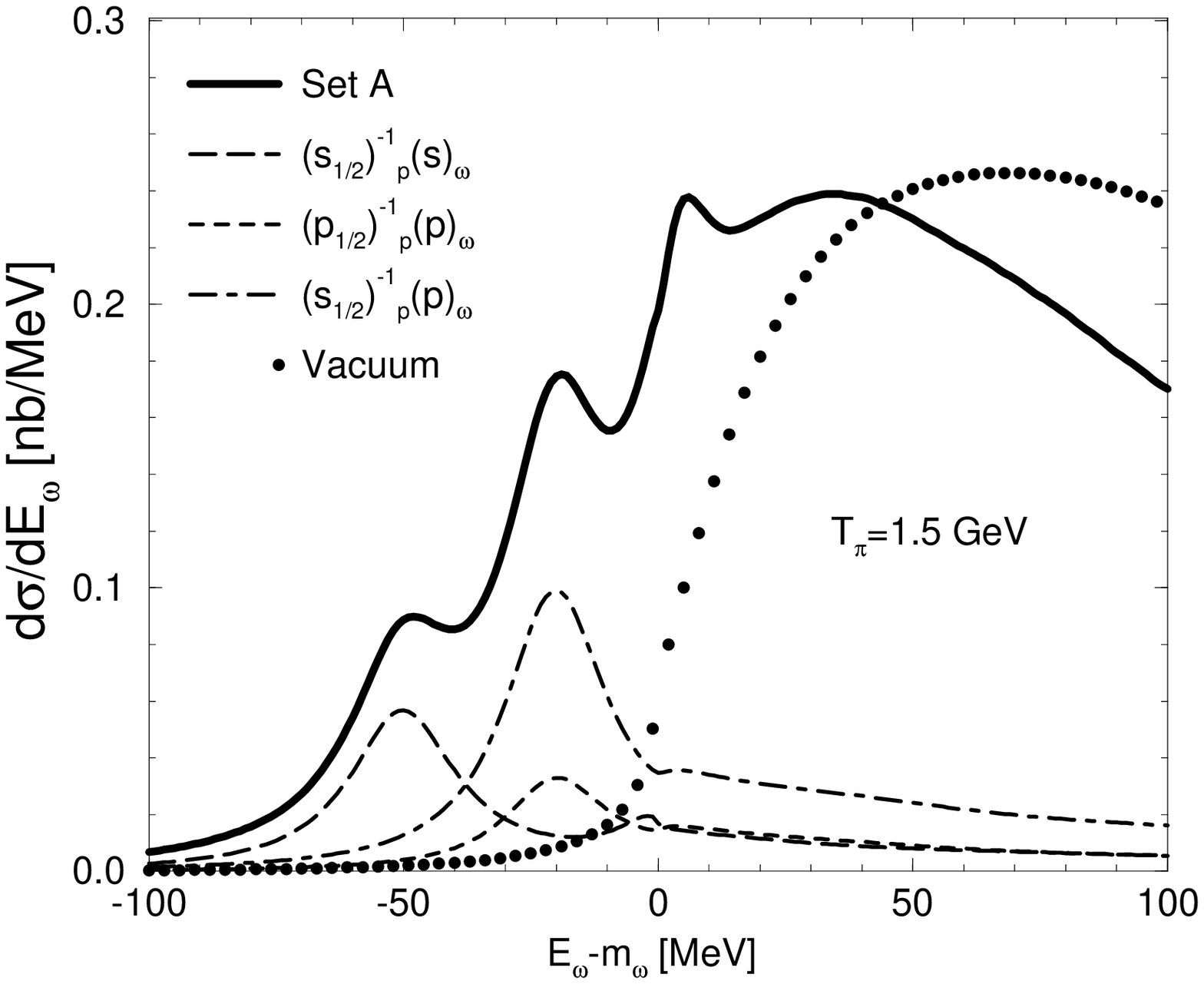,width=88mm}}
\end{picture}
\end{center}
\vspace*{-9mm}
\caption{\label{o-pi-li} Spectra of the reaction $\pi^- A\to \omega(A-1)+n$ in
case $A=7$ (Lithium), assuming that the $\omega$ meson is detected through its
decay into $e^+ e^-$. 
\protect \newline 
Solid line: result with $\omega$-nuclear potential (Set A). Dotted line ``vacuum'' result (quasi--free $\omega$ production). The separate contributions
from different bound $\omega$ meson orbitals are also shown for comparison.}
\end{figure}
where now ($-L \leq M \leq +L$):
\begin{align}
 & w_{L M}(r,\theta_n)= \\ \nonumber &  \sqrt{\pi \frac{(L-\mu)!}{(L+\mu)!}}
 \int_{-1}^{1}d\tilde{z}\, e^{-i
z(  p_\pi -p_n \cos \theta_n)} \,D(z,b)\, P_{L,\mu}(\tilde{z}) \, I_{\mu}(p_n b \sin \theta_n),
\end{align}
with $\tilde{z}=z/r$, $\mu \equiv |M|$, Legendre functions $P_{L,\mu}$ and Bessel functions $I_{\mu}$. The
 remaining integrals are solved numerically. 
 
 We show in fig.~\ref{o-pi-li} the expected results for $\pi^-+^7\!Li\to
 ^6_\omega\!He+n$ as an example, for the case in which the $\omega$ spectrum is
 detected through $e^+e^-$ pairs. We use set A for the $\omega N$ input amplitude. The strong downward shift of the
 spectral weight in the presence of the $\omega$ binding potential is quite
 apparent, and there are even indications of structures that can be associated
 with separate quasi-bound orbits of the $\omega$ meson. However, the rates are
 very low. In fact we have used the free $\omega \to e^+ e^-$ branching ratio
 for normalization, and one expects to loose about an order of
 magnitude when the in-medium branching ratio is used. 

\section{Summary and Conclusions} 
We have investigated the possible formation of quasi-bound $\omega$ meson states
in nuclei. Such states are indeed predicted to exist if the $\omega$ meson
(unlike the $\rho$ meson) survives as a quasiparticle in nuclear matter, and if
its in-medium mass is shifted downward by more than 10\% from its free mass,
$m_\omega=0.78$ GeV, as suggested by several models.

We summarize our results and conclude as follows:
\begin{itemize}
\item The complex $\omega$-nuclear potential is expected to be strongly energy
dependent, its real part changing from repulsion at low energy to attraction
($\Re U \sim -(100-150)\, \MeV \cdot \rho/\rho_0$) at $E \sim m_\omega$. This
behaviour is consistent with the QCD sum rule analysis of the first moment of
the $\omega$ meson spectral distribution in nuclear matter.

\item The estimated widths of $\omega$-nuclear states in a variety of nuclei
with $A \leq 40$ are $\Gamma \lsim 40 \, \MeV$, up to about five times larger
than the free $\omega$ decay width. Such widths may prohibit the separate
identification of $\omega$-nuclear states in distinct orbitals, but pronounced
spectral strength at excitation energies $E<m_\omega$ below the quasifree
production threshold should be visible.

\item We have examined two possible options for producing $\omega$-nuclear
states: $(d,^3\hspace{-1mm}He)$ transfer reactions and pion--induced
        formation. The $\pi+A \to \omega (A-1)+n$ process with subsequent
        detection of the $\omega$ by its decay into $e^+e^-$ is at the
        borderline of feasibility, with prohibitively low rates. On the other
        hand, missing energy spectra in $d+A \to \omega(A-1)
        +^3\hspace{-1mm}He$ reactions look more promising, with forward
        differential cross sections at the level of nb/(sr$\cdot$MeV) even for
        the lightest nuclei. Optimal kinematic conditions would be realized for
        deuteron beam kinetic energies $T_d \simeq (8$-10) GeV. Even at $T_d
        \simeq 4 $ GeV the transfered momentum is still well within the range
        of typical nuclear Fermi momenta.
\item The investigation of spectral distributions of $\omega$ mesons ``at
rest'' in normal nuclei would be complementary to the search for medium effects
on vector meson spectra in high--energy heavy--ion collisions. If such effects
are significant at high baryon densities, they should also be seen in less
extreme but much better controlled conditions as they are realized in ordinary
nuclear systems.

\end{itemize}
We would like to thank A.~Gillitzer, R.S.~Hayano, S.~Hirenzaki, P.~Kienle and
H.~Toki for fruitful discussions.

\end{document}